\documentclass[aps,prd,superscriptaddress,showkeys,showpacs,twocolumn,longbibliography,nofootinbib]{revtex4-1}  
\usepackage[utf8]{inputenc}
\usepackage{amsmath}
\usepackage{amsfonts}
\usepackage{amsthm}
\usepackage{amssymb}
\usepackage{float}
\usepackage{braket}
\usepackage{graphicx}
\usepackage{url}
\usepackage[colorlinks=true, pdfstartview=FitV, linkcolor=red, citecolor=blue, urlcolor=blue]{hyperref}
\usepackage{slashed}
\usepackage[normalem]{ulem}

\graphicspath{{plot}}

\begin{document}
\title{
Collective motion in the massive Schwinger model via Tensor Network
}
\author{Haiyang Shao}
\email[]{shy21@mails.tsinghua.edu.cn}
\affiliation{Department of Physics, Tsinghua University, Beijing 100084, China}

\author{Shile Chen}
\email[]{shile_chen@163.com}
\affiliation{Department of Physics, Tsinghua University, Beijing 100084, China}

\author{Shuzhe Shi}
\email[]{shuzhe-shi@tsinghua.edu.cn}
\affiliation{Department of Physics, Tsinghua University, Beijing 100084, China}
\affiliation{State Key Laboratory of Low-Dimensional Quantum Physics, Tsinghua University, Beijing 100084, China.}
\date{\today}

\begin{abstract}
We simulate the real-time dynamics of a massive Schwinger model using the Time-Evolving Block Decimation tensor network algorithm. Starting from a non-equilibrium initial state with localized energy excitation on top of vacuum, we track the subsequent evolution to investigate two distinct physical phenomena. First, by analyzing the system's energy-momentum tensor, we show that this system exhibits hydrodynamic behavior analogous to Bjorken flow at large coupling-to-mass ratio, a signature that diminishes as the coupling weakens, or mass increases. Second, by examining the evolution of the electric field and charge density, we observe the signal of spontaneous parity symmetry breaking phase transition in a dynamical system. The parity-restored regime is marked by ``string breaking'' and efficient charge screening, while the parity-broken regime displays stable propagation of nearly free charges and persistent electric fields connecting them.
\end{abstract}
\maketitle

\section{Introduction}
Understanding thermalization of strongly coupled quantum matter far from equilibrium is a profound challenge in modern physics. The hot and dense medium created in relativistic heavy-ion collisions, a system governed by strong interaction, is a unique example. These collisions create quark-gluon plasma, a state of matter that the color degrees of freedom become deconfined. Collective motions are found in the evolution of hot medium from the experimental results of transverse momentum distribution and multi-particle angular correlation. They are unexpectedly well-described by relativistic hydrodynamics. See e.g.~\cite{Shuryak:2014zxa, Shuryak:2003xe}.
How the quantum chromodynamics (QCD) matter rapidly develop the collective, fluid-like behavior (also referred to as ``hydrodynamization") and how they eventually convert to the color-confinement hadrons remain open questions. Owing to the non-perturbative nature of QCD, addressing these issues requires a real-time simulation of the full quantum many-body field theory.

While direct quantum simulation of the full QCD theory remains intractable due to the limitations of both classical and quantum hardwares, the Quantum Electrodynamics in one spatial dimension also known as QED$_2$ or the (massive) Schwinger model~\cite{Schwinger:1962tp}, provides a theoretical laboratory to explore these complex phenomena. Despite its simplicity, it captures essential features of QCD, including confinement, chiral symmetry breaking, and a non-trivial phase structure depending on mass-to-coupling ratio and a topological $\theta$-angle~\cite{Lowenstein:1971fc, Jayewardena:1988td, Smilga:1992hx, Adam:1993fc, Adam:1997wt, Coleman:1976uz, Adam:1995us, Adam:1996np, Adam:1996qk}. Its lower dimensionality enables quasi-exact numerical simulations with advanced techniques like tensor networks and opens a window into non-equilibrium dynamics, which remains inaccessible in full QCD.

Exact simulations of the lattice Schwinger are still limited by the hardware and constrained within at most $\sim20$ sites~\cite{Chakraborty:2020uhf, Florio:2023dke, Florio:2024aix, Ikeda:2023zil, Ikeda:2023vfk, Chen:2024pee, Ikeda:2020agk, Barata:2024bzk}.
In this work, we invoke the Matrix Product State (MPS) method~\cite{Rommer:1997zz} to represent the quantum states of interest and the Time-Evolving Block Decimation (TEBD) algorithm~\cite{Vidal:2003lvx} to simulate the real-time dynamics of the massive Schwinger model following a localized energy excitation. 
Particularly, the MPS method has been used to simulate the Schwinger model to study the vacuum mass spectrum and phase diagram~\cite{Byrnes:2002nv, Banuls:2013jaa, Banuls:2016gid, Dempsey:2023gib}, finite temperature thermodynamics~\cite{Banuls:2015sta, Banuls:2016lkq, Buyens:2016ecr}, string breaking and real-time scattering~\cite{Buyens:2013yza, Magnifico:2019kyj, Belyansky:2023rgh, Papaefstathiou:2024zsu}, collective motion created in a jet~\cite{Florio:2024aix, Florio:2025hoc, Janik:2025bbz}, thermal spectral function~\cite{Barata:2025jhd}, jet energy loss~\cite{Barata:2025hgx} and the parton distribution functions~\cite{Schneider:2024yub}.
In this work, we investigate two distinct phenomena with intrinsic connections. First, we examine how collective hydrodynamic behavior, akin to Bjorken flow, emerges and subsequently breaks down as the mass-to-coupling ratio increases. Second, we explore the dynamic signatures of the phase transition associated with spontaneous breaking of parity symmetry. We illustrate ``string breaking'' in the symmetry-restored phase versus unscreened charge separation in the symmetry-broken phase, and link these microscopic dynamics to the model's static phase diagram.

The rest of the paper is organized as follows. In Sections~\ref{sec:model} and~\ref{sec:method} we introduce the massive Schwinger model and details of tensor network methods, respectively. Section~\ref{sec:results} presents the simulation results. After a summary and outlook in Section~\ref{sec:summary}, we also provide more details regarding the convergence test in the Appendix.

\section{The Schwinger Model on a Lattice}\label{sec:model} 
Let us begin by reviewing the Schwinger model, especially the derivation of its energy-momentum tensor. We start from the continuum expressions and then derive the Hamiltonian and stress tensor operators for a discrete lattice theory, which will be used in our numerical simulations.

\subsection{Continuum Formulation and Observables}
The Lagrangian density of the massive Schwinger model with a topological $\theta$-term in $(1+1)$-dimensional Minkowski space is~\cite{Schwinger:1962tp}
\begin{align}
  \mathcal{L} = \bar{\psi}(i\overset{\leftrightarrow}{\slashed{\mathcal{D}}} - m)\psi
  - \frac{1}{4} F^{\mu\nu} F_{\mu\nu}
  + \frac{g\theta}{4\pi} \epsilon^{\mu\nu} F_{\mu\nu},
\end{align}
where the slashed covariant derivative is $\overset{\leftrightarrow}{\slashed{\mathcal{D}}} = \gamma^\mu(\overset{\leftrightarrow}{\partial}_\mu + igA_\mu)$, $A_\mu$ the $U(1)$ gauge field, $F^{\mu\nu} = \partial^\mu A^\nu - \partial^\nu A^\mu$ the corresponding electric field strength tensor, and the two-sided derivative is defined as $\bar{\psi}\gamma^\mu \overset{\leftrightarrow}{\partial}_\mu \psi = \frac{\bar{\psi}\gamma^\mu (\partial_\mu \psi) - (\partial_\mu \bar{\psi})\gamma^\mu \psi}{2}$. In $(1+1)$ dimensions, the fermion field $\psi$ is a two-component Dirac spinor, and $m$ and $g$---respectively being the fermion mass and gauge coupling---both carry an energy unit.
We take the metric convention $g_{\mu\nu} = \mathrm{diag}(1, -1)$ throughout the paper.

The $\theta$-term can be reformulated into the mass term by a chiral transformation: $\psi \to e^{i\gamma_5\theta/2}\psi$, $\bar{\psi} \to \bar{\psi} e^{i\gamma_5\theta/2}$. Up to a boundary term, the Lagrangian then becomes
\begin{align}
  \label{eq:schwinger_compact}
  \mathcal{L} =
  \bar{\psi}(i\overset{\leftrightarrow}{\slashed{\mathcal{D}}}
  - m\, e^{i\gamma_5\theta})\psi
  - \frac{1}{4} F^{\mu\nu} F_{\mu\nu}.
\end{align}
The model exhibits rich phenomena such as confinement and deconfinement, with the phase structure determined by the ratio $m/g$ and the value of $\theta$.

To derive the Hamiltonian, one may start from the canonical momenta conjugate to $A_\mu$ and $\psi$ fields,
\begin{align}
\begin{split} 
\mathcal{E}_0 & \equiv \frac{\delta \mathcal{L}}{\delta \dot{A}_0}=0, \\
\mathcal{E} & \equiv \frac{\delta \mathcal{L}}{\delta \dot{A}_1}=\partial_t A_1-\partial_z A_0, \\
\pi_\psi & \equiv \frac{\delta \mathcal{L}}{\delta\left(\partial_t \psi\right)}=\frac{i}{2} \bar{\psi} \gamma^0,\\
\pi_{\bar{\psi}} & \equiv \frac{\delta \mathcal{L}}{\delta\left(\partial_t \bar{\psi}\right)}=-\frac{i}{2} \gamma^0 \psi,
\end{split} 
\end{align}
and then perform the Legendre transformation,
\begin{equation}
\begin{aligned}
\mathcal{H} 
= \;& \frac{\mathcal{E}^2}{2}
    +\bar{\psi}(-i\gamma^1 \overset{\leftrightarrow}{\partial_z} +g\gamma^1 A_1 + me^{i\gamma_5 \theta})\psi\\
& +\left(\partial_z \mathcal{E}-g \bar{\psi} \gamma^0 \psi\right) \,g\,A_0\,.
\end{aligned}\label{eq:Hamiltonian_cont}
\end{equation}
Particularly, $\mathcal{E}=F_{01}$ is the electric field
in the Schr\"odinger picture, and the field dynamics are defined by the commutation and anticommutation relations: $\big[A_1(z), \mathcal{E}\left(z'\right)\big] = i\,\delta(z-z')$ and $\big\{\psi_a(z), \pi_\psi^b(z')\big\} = i\,\delta_a^b \delta\left(z-z^{\prime}\right)$.

While the Lagrangian $\mathcal{L}$ is explicitly invariant under a gauge transformation,
\begin{equation}
    A_\mu \rightarrow A_\mu-g^{-1} \partial_\mu \varphi_\mathrm{g}, \qquad \psi \rightarrow e^{i \varphi_\mathrm{g}} \psi,\label{eq:gauge_transformation}
\end{equation}
the Hamiltonian transforms as $\mathcal{H} \rightarrow \mathcal{H}-\left(\partial_z \mathcal{E}-g \bar{\psi} \gamma^0 \psi\right) \partial_t \varphi_\mathrm{g}$. Thus, the Gauss' law,
\begin{equation}
    \partial_z \mathcal{E} = g\, \bar{\psi} \gamma^0 \psi\,,
\end{equation}
must be strictly satisfied for all physical states, to ensure the gauge invariance of the Hamiltonian, and the second line in Eq.~\eqref{eq:Hamiltonian_cont} shall be omitted.

This constraint means the electric field $\mathcal{E}(z)$ is not an independent degree of freedom in 1+1 D. Instead, it is determined by the integrated charge distribution from a reference point $z_0$:
\begin{equation}
    \mathcal{E}(z)=\mathcal{E}_{\text{bnd}}+g \int_{z_0}^z \bar{\psi}\left(z^{\prime}\right) \gamma^0 \psi\left(z^{\prime}\right) \mathrm{d} z^{\prime},
\end{equation}
where $\mathcal{E}_{\text{bnd}}$ is the boundary field at $z_0$.

When considering collective phenomena near thermal equilibrium, we always focus on the Noether currents associated with symmetries, i.e., charge current and energy-momentum stress tensor in this theory.
For the Schwinger model on a finite interval $z \in[0, L]$, we impose open boundary conditions where the fermion and electric fields vanish: $\psi(L)=\psi(0)=0$ and $\mathcal{E}(L)=\mathcal{E}(0)=\mathcal{E}_{\text{bnd}}=0$. Substituting these conditions into the expression for $\mathcal{E}(L)$ forces the total charge of the system to be zero [note that $\hat{j}^t(z) \equiv \bar{\psi}(z) \gamma^0 \psi(z)$ is the local charge density]
\begin{equation}
    0=\int_0^L \bar{\psi}(z) \gamma^0 \psi(z) \mathrm{d} z\,.
\end{equation}
Thus, physical states under these boundary conditions must be charge-neutral. 

The canonical stress tensor can be derived from two approaches. First, it is the Noether current associated with spacetime translations,
\begin{align}
    \hat{T}^{\mu\nu}_\mathrm{can} = \frac{\delta \mathcal{L}}{\delta (\partial_\mu \hat\varphi_j)} \partial^\nu \hat\varphi_j - g^{\mu\nu} \mathcal{L},
\end{align}
where \(\hat\varphi_j\) represents the fields (\(\psi\), \(\bar{\psi}\), \(A_\lambda\)). Throughout the manuscript, we address the stress tensor operators with $\hat{T}$, to distinguish them from corresponding expectation values. Applying this to the Schwinger Lagrangian~\eqref{eq:schwinger_compact} results in
\begin{align}
    \hat{T}^{\mu\nu}_\mathrm{can} = i\,\bar{\psi} \gamma^\mu \overset{\leftrightarrow}{\partial^\nu} \psi - F^{\mu\rho} \partial^\nu A_\rho - g^{\mu\nu} \mathcal{L},
\end{align}
which is neither symmetric nor gauge-invariant. This shortcoming can be fixed by applying the Belinfante--Rosenfeld procedure~\cite{1939Phy.....6..887B, Rosenfeld1940}. This procedure corrects the canonical tensor by adding into the stress tensor the divergence of an antisymmetric tensor, $\partial_\rho \chi^{\mu\rho\nu}$, derived from the fields' spin density tensor via $\chi^{\mu\rho\nu} = -\frac{1}{2} (\mathcal{S}^{\mu\rho\nu} - \mathcal{S}^{\nu\mu\rho} + \mathcal{S}^{\rho\nu\mu})$. We compute the latter as the Noether currents associated with Lorentz transformations,
$\mathcal{S}^{\mu\rho\nu} = \frac{1}{2} \bar{\psi} (\gamma^\mu \Sigma^{\rho\nu} + \Sigma^{\rho\nu} \gamma^\mu) \psi -F^{\mu\rho} A^\nu + F^{\mu\nu} A^\rho$, with $\Sigma^{\rho\nu} = i\,[\gamma^\rho, \gamma^\nu]/4$. Applying Dirac and Maxwell equations, we obtain the correction term
\begin{align}
\begin{split}
    \partial_\rho \chi^{\mu\rho\nu} =& 
    \frac{i}{2} (\bar{\psi} \gamma^\nu \overset{\leftrightarrow}{\mathcal{D}^\mu} \psi - \bar{\psi} \gamma^\mu \overset{\leftrightarrow}{\mathcal{D}^\nu} \psi)
        + F^{\mu\rho} \partial_\rho A^\nu,
\end{split}
\end{align}
as well as the Belinfante stress tensor,
\begin{align}
\begin{split}
\hat{T}^{\mu\nu} \equiv\; & 
    \hat{T}^{\mu\nu}_\mathrm{can} + \partial_\rho \chi^{\mu\rho\nu}\\
    =\; &\frac{i}{2}\bar\psi \big(\gamma^\mu \overset{\leftrightarrow}{\mathcal{D}^\nu} + \gamma^\nu \overset{\leftrightarrow}{\mathcal{D}^\mu} \big) \psi
    - F^{\mu\alpha}  {F^\nu}_\alpha - g^{\mu\nu} \mathcal{L}\,,
    \end{split}\label{eq.stress_tensor}
\end{align}
which is symmetric ($\hat{T}^{\mu\nu} = \hat{T}^{\nu\mu}$) and gauge-invariant. 
One can show that the Belinfante stress tensor is equivalent to the one obtained in the second approach---functionally differentiating the action with respect to the spacetime metric, $\hat{T}^{\mu\nu} = \frac{2}{\sqrt{-g}} \frac{\delta}{\delta g_{\mu\nu}} \int \mathcal{L} \,\mathrm{d}t\,\mathrm{d}z$.

In the Schr\"odinger picture, different components of the stress tensor read,
\begin{align}
\begin{split}
\hat{T}^{tt} 
=\;& 
    \frac{\mathcal{E}^2}{2} 
    - i \bar{\psi} \gamma^1 \overset{\leftrightarrow}{\partial_z}\psi
    + \bar{\psi}(
     g\gamma^1 A_1 + m\,e^{i\gamma_5\theta})\psi\,,\\
\hat{T}^{zz} 
=\;&
    -\frac{\mathcal{E}^2}{2} 
    - i \bar{\psi} \gamma^1 \overset{\leftrightarrow}{\partial_z}\psi
    + g \bar{\psi} \gamma^1 A_1\psi
\,,\\
\hat{T}^{zt} 
=\;&
\hat{T}^{tz}
=
     -i\psi^\dagger \overset{\leftrightarrow}{\partial_z}\psi 
     + g\,A_1\, \psi^\dagger \psi.
\end{split}\label{eq.stress_tensor_3}
\end{align}
We note that energy-momentum conservation is ensured explicitly at the operator level, $[\hat{T}^{t\nu},H] = -i\,\partial_z \hat{T}^{z\nu}$.

To study the hydrodynamic behavior of the Schwinger model, we decompose the stress tensor into its hydrodynamic components~\cite{1959flme.book.....L},
\begin{align}  
    T^{\mu\nu} = (\varepsilon + P + \Pi) u^\mu u^\nu - (P+\Pi)\, g^{\mu\nu} + \pi^{\mu\nu},  
\end{align}  
where $\varepsilon$ is the energy density, $P$ the pressure, $u^\mu$ the flow velocity, $\Pi$ the bulk pressure, and $\pi^{\mu\nu}$ the shear stress tensor. We have taken the Landau frame~\cite{1959flme.book.....L}, which defines the velocity to be comoving with the energy,
\begin{equation}
    {T^{\mu}}_{\nu}\, u^\nu = \varepsilon\, u^\mu\,.
    \label{eq:landau_frame}
\end{equation}

Noting that $\pi^{\mu\nu}$ is defined to be traceless and orthogonal to the flow velocity, it vanishes in $(1+1)$ dimensions. The pressure and the energy density are related by the thermodynamic equation of state---they are expectations of the $\hat{T}^{zz}$ and $\hat{T}^{tt}$ operators in a thermal ensemble, respectively. $\Pi$ measures the difference between the effective pressure and the thermal one and characterizes the degree of off-equilibrium.

In $(1+1)$D, the Landau frame eigenvalue equation~\eqref{eq:landau_frame} has two simple solutions,
\begin{align}  
\left\{
\begin{array}{l}
    \varepsilon = T_- + R, \quad 
    v^z  = \frac{T^{tz}}{T_+ + R}, \quad  
    P+\Pi  = R - T_-
;\\
    \varepsilon = T_- - R, \quad
    v^z = \frac{T^{tz}}{T_+ - R}, \quad
    P+\Pi = -R - T_-\,,
\end{array}
\right.
\label{eq:hydro_variables}
\end{align}
where
\begin{align}
\begin{split}  
    T_\pm \equiv \frac{T^{tt} \pm T^{zz}}{2}, \qquad
    R \equiv \sqrt{(T_+)^2 - (T^{tz})^2}.  
\end{split}  
\end{align}
One should take the solution with a time-like flow velocity, $v^z \equiv \frac{u^z}{u^t} \in (-1,1)$.

\subsection{Lattice Formulation}
In numerical simulations, one needs to define the space-dependent operators on a discrete lattice. The spatial coordinate $z$ is replaced by lattice sites $z_n = n\, a$, with lattice spacing $a$ and $n = 1, \cdots, N$. The gauge field is represented by link operators scaled to be unitless,  
\begin{align}  
L_n \equiv g^{-1} \mathcal{E}(z_n), \qquad
\phi_n \equiv a\,g\,A_1(z_n),  
\end{align} 
with the commutation relation:  
\begin{align}
    [\phi_n, L_m] = i\, \delta_{nm}, \qquad [e^{i\phi_n}, L_m] = -e^{i\phi_n}\, \delta_{nm}\,.
\end{align}  

Let $|l\rangle_n$ denote an eigenstate of $L_n$ at site $n$ with eigenvalue $l$, i.e., $L_n |l\rangle_n = l |l\rangle_n$. From the commutation relation, we can see that $e^{i\phi_n}$ acts as a raising operator on these states, 
\begin{align}
\begin{split}
    L_n (e^{i\phi_n}|l\rangle_n) = (e^{i\phi_n}L_n + e^{i\phi_n})|l\rangle_n 
    = (l+1)(e^{i\phi_n}|l\rangle_n).  
\end{split}
\end{align} 

In our numerical implementation, we work in the eigenbasis of the electric field operator $L_n$,
\begin{align}
    \begin{split}
        L_n = \sum_{l=-\infty}^{\infty} l\, |l\rangle_n \langle l|_n, \qquad
        e^{i\phi_n} = \sum_{l=-\infty}^{\infty} |l+1\rangle_n \langle l|_n.
    \end{split}
\end{align}
Note that the spectrum of $L_n$ is, in principle, infinite ($l \in \mathbb{Z}$), for practical computations we must truncate the Hilbert space to a finite dimension. We introduce a cutoff $L_{\text{max}}$ such that the allowed eigenvalues are restricted to $l \in \{-L_{\text{max}}, -L_{\text{max}}+1, \dots, L_{\text{max}}\}$. This truncation introduces an error by imposing the unitarity of the raising operator $e^{i\phi_n}$---when the operator acts on the highest state $|L_{\text{max}}\rangle_n$, it wraps around to the lowest state:  
\begin{align}  
    e^{i\phi_n}|L_{\text{max}}\rangle_n \rightarrow |-L_{\text{max}}\rangle_n.  
\end{align}  

To illustrate the structure of these operators in the truncated space, we consider the minimal non-trivial case with a cutoff $L_{\text{max}}=1$. For $L_{\text{max}}=1$, the local Hilbert space at each site is three-dimensional, spanned by the basis states $\{|-1\rangle_n, |0\rangle_n, |1\rangle_n\}$. In this basis, the operators $L_n$ and $e^{i\phi_n}$ are represented by the following $3 \times 3$ matrices:  
\begin{align}  
    L_n =   
    \begin{pmatrix}  
        -1 & 0 & 0 \\
        0 & 0 & 0 \\
        0 & 0 & 1  
    \end{pmatrix},  
    \qquad  
    e^{i\phi_n} =   
    \begin{pmatrix}  
        0 & 0 & 1 \\
        1 & 0 & 0 \\
        0 & 1 & 0  
    \end{pmatrix}.  
\end{align}
The matrix for $e^{i\phi_n}$ explicitly performs the shift operation: it maps $|-1\rangle \to |0\rangle$, $|0\rangle \to |1\rangle$, and ensures the unitarity by mapping $|1\rangle \to |-1\rangle$. Given this subtlety, in the practical numerical calculations presented later in this work, we will employ a sufficiently large cutoff to ensure convergence of observables of interest.

\vspace{3mm}
The fermionic fields are discretized using the Kogut--Susskind staggered fermion formalism~\cite{Kogut:1974ag, Banks:1975gq}, 
\begin{equation}
    \chi_{2 n}=a^{\frac{1}{2}} \psi_{\uparrow}\left(z_{2 n}\right), \quad \chi_{2 n+1}=a^{\frac{1}{2}} \psi_{\downarrow}\left(z_{2 n+1}\right).
\end{equation}
They follow the anticommutation relations
\begin{equation}
    \left\{\chi_n^{\dagger}, \chi_m\right\}=\delta_{n m}, \quad\left\{\chi_n^{\dagger}, \chi_m^{\dagger}\right\}=\left\{\chi_n, \chi_m\right\}=0\,.
\end{equation}

We can map them to spin operators via the Jordan--Wigner transformation~\cite{Jordan:1928wi},
\begin{align}  
\begin{split} 
& \chi_n=S_n^- \prod_{m=1}^{n-1}\left(-i Z_m\right), \qquad
 \chi_n^{\dagger}=S_n^+ \prod_{m=1}^{n-1}\left(i Z_m\right),
\end{split}  
\end{align}
where we have used the notation $S_n^{\pm} \equiv\left(\prod_{j=1}^{n-1} \otimes I\right) \otimes \frac{1}{2}(X_n \pm i Y_n) \otimes$ $\left(\prod_{j=n+1}^N \otimes I\right)$ and likewise for $Z_n$. 

The resulting lattice Hamiltonian under open boundary conditions is:  
\begin{align}  
\begin{split}  
    H =\;& \frac{ag^2}{2}\sum_{n=1}^{N-1}L_n^2 + \frac{1}{2a}\sum_{n=1}^{N-1}(S_n^+  e^{i\phi_n}S_{n+1}^- + S_{n+1}^+  e^{-i\phi_n}S_{n}^-)\\
    &+ \frac{m\sin\theta}{2}\sum_{n=1}^{N-1}(-1)^{n+1}(S_n^+  e^{i\phi_n}S_{n+1}^- + S_{n+1}^+  e^{-i\phi_n}S_{n}^-)\\
    &+ \frac{m\cos\theta}{2}\sum_{n=1}^{N}(-1)^n Z_n\,.\\
\end{split}  
\end{align}
It is worth mentioning that in a lattice theory with fermion mass $m_{\text{lat}}$ corresponds to a continuous theory with effective fermion mass $m_{\text{con}} = m_{\text{lat}} -\frac{g^2a}{8}$~\cite{Dempsey:2022nys}.

For better analysis of the spatial dependence of the hydrodynamic variables, we take the follows lattice expression of local charge and energy-momentum densities, which are all symmetric with respect to a spatial reflection centered at $z_n$,
\begin{align} 
\begin{split}  
\hat{j}^t_n =\;& 
    \frac{Z_n+(-1)^n}{2 a}\,,\\
\hat{T}^{tt}_{n} =\;& 
    \left(\frac{1}{4a^2}+ \frac{(-1)^{n+1}m\sin\theta}{4a}\right)\left(S_n^{+} e^{i\phi_n} S_{n+1}^{-}  \right.
    \\
    &  \left. + S_{n-1}^{+} e^{i\phi_{n-1}} S_n^{-} + S_{n+1}^{+} e^{-i\phi_n} S_{n}^{-} + S_{n}^{-} e^{-i\phi_{n-1}} S_n^{+} \right)\\
    &+ \frac{g^2}{4}(L_{n-1}^2 + L_{n}^2) + \frac{(-1)^n m\cos\theta}{2a}(1+Z_n),\\
\hat{T}^{zz}_{n} =\;& 
    \frac{1}{4a^2} \left(S_n^{+} e^{i\phi_n} S_{n+1}^{-} + S_{n-1}^{+} e^{i\phi_{n-1}} S_n^{-}\right.\\
    &\left.+S_{n+1}^{+} e^{-i\phi_n} S_{n}^{-} + S_{n}^{+} e^{-i\phi_{n-1}} S_{n-1}^{-}\right) \\
    &- \frac{g^2}{4}(L_{n-1}^2 + L_{n}^2),\\
\hat{T}^{tz}_{n} =\;& \hat{T}^{zt}_{n} = 
    \frac{-i}{4a^2}\left(S_{n-1}^{+}Z_nS_{n+1}^{-}e^{i\phi_n}e^{i\phi_{n-1}}\right.\\
    &\left. \qquad \qquad - S_{n+1}^{+}Z_nS_{n-1}^{-}e^{-i\phi_n}e^{-i\phi_{n-1}}\right).  
\end{split}  
\end{align}  

\section{Tensor Network Theory}\label{sec:method}
Simulating the quantum dynamics of many-body systems from first principles is a fundamental computational challenge. While the lattice Schwinger model can be solved exactly on a classical computer~\cite{Florio:2023dke, Florio:2024aix, Ikeda:2023zil, Ikeda:2023vfk, Chen:2024pee, Ikeda:2020agk, Barata:2024bzk}, the method is limited to small systems. The difficulty lies in the size of the Hilbert space. A general pure state $|\Psi\rangle$ on an $N$-site lattice is a vector in a space of dimension $d^N$,
\begin{equation}
    |\Psi\rangle = \sum_{\kappa_1, \ldots, \kappa_N} C_{\kappa_1, \ldots, \kappa_N} |\kappa_1, \ldots, \kappa_N\rangle,
    \label{eq:fullstate}
\end{equation}
where $|\kappa_i \rangle$ represents the quantum state on the $i$-th site and it has $d$ different choices.
Storing and manipulating the $d^N$ coefficients $C_{\kappa_1, \ldots, \kappa_N}$ become computationally intractable for even modestly-sized systems (e.g., $N = 24$ for $d=2$), which is a problem known as the ``curse of dimensionality.'' As studying the hydrodynamic behavior relies on the space-time profile of the medium and needs as many grids as possible, this exponential scaling makes exact methods unsuitable.

Fortunately, the states relevant to many physical phenomena are not generic, highly-entangled vectors but instead occupy a tiny, structured corner of the Hilbert space. This structure is governed by the principle of limited entanglement. For ground states, it has been rigorously proven that one-dimensional, local, and gapped Hamiltonians obey a strict ``area law''~\cite{Hastings:2007iok}---that is, the entanglement entropy of a subregion scales with its boundary area, which for a contiguous interval in 1D is constant and independent of the interval's length.  Furthermore, for dynamics, a state prepared with low entanglement (such as a ground state or a low-lying excitation) and evolved under a local Hamiltonian will only generate entanglement at a finite rate~\cite{Bravyi:2006zz}. Consequently, for the finite evolution times considered in simulations, the state remains limitedly entangled. The tensor network framework, and specifically the Matrix Product State (MPS) ansatz~\cite{Rommer:1997zz}, is a powerful representation built precisely to exploit this low-entanglement structure, allowing for efficient simulations on large lattices.

In this section, we detail the tensor network methodologies employed for our investigation. We begin by outlining the procedure for simulating the non-equilibrium, real-time dynamics using an MPS. Subsequently, we describe the calculation of the equation of state (EoS) via the purification method, which establishes the equilibrium thermodynamic properties required for the analysis of the system's dynamic behavior.

\subsection{Real time evolution of pure-state vector}
The MPS ansatz parameterizes the $d^N$ coefficients of the state vector in Eq.~\eqref{eq:fullstate} as a contraction of $N$ local, rank-3 tensors:
\begin{equation}
    C_{\kappa_1, \ldots, \kappa_N} = \sum_{\alpha_1, \ldots, \alpha_{N-1}} \left[A^{\kappa_1}_1\right]_{\alpha_0, \alpha_1} \left[A^{\kappa_2}_2\right]_{\alpha_1, \alpha_2} \cdots \left[A^{\kappa_N}_N\right]_{\alpha_{N-1}, \alpha_N}.
\end{equation}
Here, each $A_n$ is a site-specific tensor with a ``physical index'' $\kappa_n$ and two ``virtual indices'' $\alpha_{n-1}, \alpha_n$ that link to adjacent sites. For open boundary conditions, the external indices $\alpha_0$ and $\alpha_N$ are trivial (dimension 1). For our study of the Schwinger model, the local basis state is a composite object $|\kappa_n\rangle = |s_n, l_n\rangle_n$, where $|s_n\rangle$ is the eigenbasis for fermion operator $Z_n$ (with $s_n=\pm 1$) and $|l_n\rangle$ is the eigenbasis for the electric field $L_n$. With the truncation at $L_{\text{max}}$ as defined previously, the total local dimension is $d = d_s \, d_l = 4L_{\text{max}} + 2$. The maximum dimension of the internal virtual indices is called the bond dimension and denoted as $D$. This parameter controls the number of parameters, which scales polynomially as $\mathcal{O}(N d D^2)$, and more importantly, it dictates the maximum amount of bipartite entanglement the MPS can capture: the von Neumann entropy $S$ for any cut is bounded by $S \le \log D$. The success of the simulation thus relies on using a bond dimension $D$ large enough to accommodate the entanglement present in the system.

\begin{figure}[htbp]
    \centering
    \includegraphics[width=0.3\textwidth]{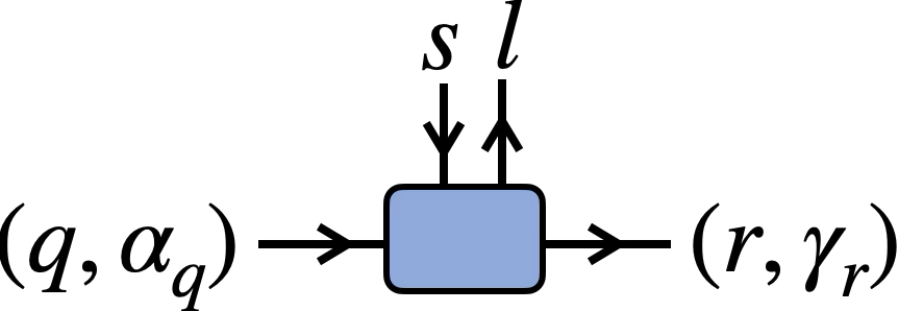}
    \caption{Schematic of the $U(1)$ gauge-invariant tensor for the pure state. The tensor receives the state of the electric field from the left link, encoded by the virtual index $q$. It combines this with the local matter field state (spin $s$) to determine the state of the outgoing electric field, $l$. Finally, a new virtual index encoding this outgoing field value, $r$, is passed to the right. The tensor's structure enforces two conditions: Gauss's law is obeyed ($l$ is fixed by $q$ and $s$), and the virtual and physical representations are consistent ($r=l$).
    }
    \label{fig:GIMPS}
\end{figure}

In our lattice Schwinger model, Gauss' law, $\hat{G}_n \equiv L_n - L_{n-1} - \frac{Z_n + (-1)^n}{2}=0$, needs to be fulfilled to ensure the gauge symmetry. To satisfy this constraint, we employ a $U(1)$ gauge-invariant MPS \cite{2011PhRvB..83k5125S, Buyens:2013yza} that constrains the Hilbert space to the physical subspace. For clarity in this gauge-invariant construction, we now label the left and right virtual indices of tensor $A_n$ as $\alpha$ and $\gamma$, which correspond to $\alpha_{n-1}$ and $\alpha_n$ in the general notation above, respectively. In the gauge-invariant framework, these indices are given a multi-index structure, $\alpha \rightarrow (q,\alpha_q)$ and $\gamma \rightarrow (r,\gamma_r)$, where $q$ and $r$ label the virtual charge sectors, representing the electric fields on the left and right links, respectively. The tensors decompose into a direct sum, $A_n =\oplus_{q,s,l,r} \left[A^{s, l}_n\right]_{\left(q, \alpha_q\right) ;\left(r, \gamma_r\right)}$, with the non-zero blocks defined as:
\begin{equation}
    \left[A^{s, l}_n\right]_{\left(q, \alpha_q\right) ;\left(r, \gamma_r\right)}=\left[a^{q, s, l, r}_n\right]_{\alpha_q, \gamma_r} \delta_{l, r} \,\delta_{q+\frac{s+(-1)^n}{2}, l}  \,,
\end{equation}
where $[a^{q, s, l, r}_n] \in \mathbb{C}^{D_q \times D_r}$ and the total bond dimension of the MPS is $D=\sum_{q \in \mathbb{Z}} D_q$. The Kronecker deltas explicitly enforce Gauss's law at the tensor level. The first, $\delta_{l,r}$, ensures the right virtual charge $r$ matches the physical electric field $l$ on the outgoing link. The second, $\delta_{q+\frac{s+(-1)^n}{2}, l}$, is Gauss's law itself: the outgoing field $l$ is fixed by the incoming field $q$ from the left link and the local charge $(s+(-1)^n)/2$ at site $n$. Figure~\ref{fig:GIMPS} provides a schematic of this gauge-invariant tensor structure.

The real time simulation of non-equilibrium dynamics is a three-step process. First, we prepare the system's ground state, $|\Psi_{GS}\rangle$, via imaginary-time evolution using the finite-site Time-Evolving Block Decimation (TEBD) algorithm \cite{Vidal:2003lvx} with open boundary conditions:
\begin{equation}
    |\Psi_{GS}\rangle = \lim_{\tau \to \infty} \mathcal{N}_{\mathrm{norm}}(\tau) e^{-\tau \hat{H}} |\Psi\rangle,
\end{equation}
where $|\Psi\rangle$ is a random initial state, $\hat{H}$ is the system Hamiltonian, and $\mathcal{N}_{\mathrm{norm}}$ the $\tau$-dependent normalization factor.

Second, we prepare the initial state $|\Psi(t=0)\rangle$ by applying a localized excitation to the ground state, to mimic the initial state of high-energy collisions which excites a small region with high-energy density on top of the vacuum. We generate the excitation by the unitary operator $U(\varphi, \hat{\nu}_n) = e^{-i \varphi \hat{\nu}_n}$, where $\hat{\nu}_n$ is the local scalar condensate density,
\begin{equation}
    \hat{\nu}_n = \bar{\psi} \psi = \frac{(-1)^n}{2}(1+\hat{Z}_n).
\end{equation}
This operator preserves gauge symmetry, satisfying $[\hat{\nu}_n, \hat{G}_{n'}] = 0$ for all $n, n'$. We set the excitation strength $\varphi = \pi$ and apply the operator at sites $n=50$ and $n=51$:
\begin{equation}
    |\Psi(t=0)\rangle = U(\pi, \hat{\nu}_{50}) \, U(\pi, \hat{\nu}_{51}) \, |\Psi_{GS}\rangle.
    \label{eq:initial_state}
\end{equation}

Third, we simulate the real-time evolution of this initial state:  
\begin{equation}
    |\Psi(t)\rangle = e^{-i t \hat{H}}|\Psi(0)\rangle,
\end{equation}
again using the TEBD algorithm. At each time step, we compute the expectation values of relevant local operators---including the stress tensor ($T^{\mu\nu}$), electric field ($\mathcal{E}$), and charge density ($j^t$)---which provide the raw data for our analysis:
\begin{align}
    \begin{split}
    \langle \hat{T}^{\mu\nu}(na, t) \rangle =& \langle\Psi(t)|\hat{T}^{\mu\nu}_n|\Psi(t)\rangle, \\
    \langle \hat{\mathcal{E}}(na, t) \rangle =& g\langle\Psi(t)|L_n|\Psi(t)\rangle, \\
    \langle \hat{j}^t(na, t) \rangle =& \frac{\langle\Psi(t)|Z_n|\Psi(t)\rangle+(-1)^n}{2 a}.
    \end{split}
\end{align}

As derived in the preceding section, a tensor decomposition of the measured expectation values $\langle \hat{T}^{\mu\nu} \rangle$ yields the energy density $\varepsilon$, fluid velocity $v^z$, and the effective pressure $P+\Pi$.

\subsection{Thermodynamic properties with density matrix}
The analysis of the time-dependent stress-energy tensor $\langle \hat{T}^{\mu\nu} \rangle$ yields the effective $P+\Pi$. We need to determine the thermodynamic pressure $P(\varepsilon)$ for the same system, so that the bulk viscous pressure $\Pi$ can also be computed.

This requires computing the thermal density matrix $\hat{\rho}$ at various inverse temperatures $\beta=1/T$. A direct Matrix Product Operator (MPO) representation of the density matrix may fail to guarantee positivity due to truncation errors in the tensor network. To overcome this, we employ the purification method \cite{Verstraete:2004gdw}, which reformulates the mixed thermal state as a pure state in an enlarged space, formally known as the Thermofield Double (TFD) state. Instead of representing $\hat{\rho}(\beta)$ directly, we construct this pure state $|\Phi(\beta)\rangle$ in $\mathcal{H} \otimes \mathcal{H}^a$, where $\mathcal{H}$ is the original physical Hilbert space and $\mathcal{H}^a$ is a fictitious auxiliary copy (ancilla) that mirrors $\mathcal{H}$ to encode correlations ensuring positivity.

The procedure begins at infinite temperature ($\beta=0$), where the density matrix is maximally mixed, $\hat{\rho}(0) \propto \hat{I}$. Its purification, $|\Phi(0)\rangle$, is a maximally entangled state between the physical and auxiliary spaces:
\begin{equation}
    |\Phi(0)\rangle = \frac{1}{\sqrt{\mathcal{C}}}\sum_{\boldsymbol{\kappa}, \boldsymbol{\kappa}^a} \delta_{\boldsymbol{\kappa}, \boldsymbol{\kappa}^a}\left|\kappa_1, \kappa_1^a, \ldots, \kappa_{N}, \kappa_{N}^a\right\rangle,
\end{equation}
where the sum is over the complete orthonormal basis $|\boldsymbol{\kappa}\rangle$ of the physical space $\mathcal{H}$, and $|\boldsymbol{\kappa}^a\rangle$ of the auxiliary space $\mathcal{H}^a$.  The normalization factor $\mathcal{C}$ is the total dimension of the physical Hilbert space. Tracing out the auxiliary space correctly recovers the maximally mixed state.

To obtain the thermal state at a finite temperature ($\beta > 0$), we apply an imaginary-time evolution operator to this infinite-temperature state:
\begin{equation}
    |\Phi(\beta)\rangle = \mathcal{N} \, e^{-\beta \hat{H}/2} \otimes \hat{I}^a |\Phi(0)\rangle,
\end{equation}
where $\mathcal{N}$ is a normalization constant, and $\hat{I}^a$ is the identity operator on the auxiliary space. The evolution operator $e^{-\beta \hat{H}/2}\otimes \hat{I}^a$ acts only on the physical Hilbert space $\mathcal{H}$. We efficiently perform this operation using the TEBD algorithm.

Once the TFD state $|\Phi(\beta)\rangle$ has been prepared, the physical density matrix is recovered by tracing out the auxiliary degrees of freedom:
\begin{equation}
    \hat{\rho}(\beta) = \mathrm{Tr}_{\mathcal{H}^a}\left(|\Phi(\beta)\rangle \langle \Phi(\beta)|\right).
\end{equation}
By construction, any state purified in this manner yields a valid, positive semi-definite density matrix. Consequently, the thermal expectation value of a physical operator $\hat{O}$, which acts only on $\mathcal{H}$, is computed as a standard expectation value with respect to the pure state $|\Phi(\beta)\rangle$:
\begin{equation}
    \langle \hat{O} \rangle_\beta \equiv \mathrm{Tr}(\hat{\rho}(\beta) \hat{O}) = \langle\Phi(\beta)|\hat{O} \otimes \hat{I}^a|\Phi(\beta)\rangle.
\end{equation}

Numerically, we represent the TFD state $|\Phi(\beta)\rangle$ as an MPS built from local tensors $B_n$. Each tensor has its physical dimension expanded to span both the physical basis $|\kappa_n\rangle=|s_n, l_n\rangle_n$ and the auxiliary basis $|\kappa_n^a\rangle=|s_n^a, l_n^a\rangle_n$. The state is then constructed as:
\begin{align}
    \begin{split}
        |\Phi(\beta)\rangle = &\sum_{\boldsymbol{\kappa}, \boldsymbol{\kappa}^a} \sum_{\boldsymbol{\alpha}}  \left[B^{\kappa_1, \kappa_1^a}_1\right]_{\alpha_0, \alpha_1}\cdots \left[B^{\kappa_N, \kappa_N^a}_N\right]_{\alpha_{N-1}, \alpha_N} \\
        &\left|\kappa_1, \kappa_1^a, \ldots, \kappa_{N}, \kappa_{N}^a\right\rangle,
    \end{split}
\end{align}
where we assume open boundary conditions, setting the boundary virtual indices $\alpha_0$ and $\alpha_N$ to be of dimension 1. 

\begin{figure}[htbp]
    \centering
    \includegraphics[width=0.3\textwidth]{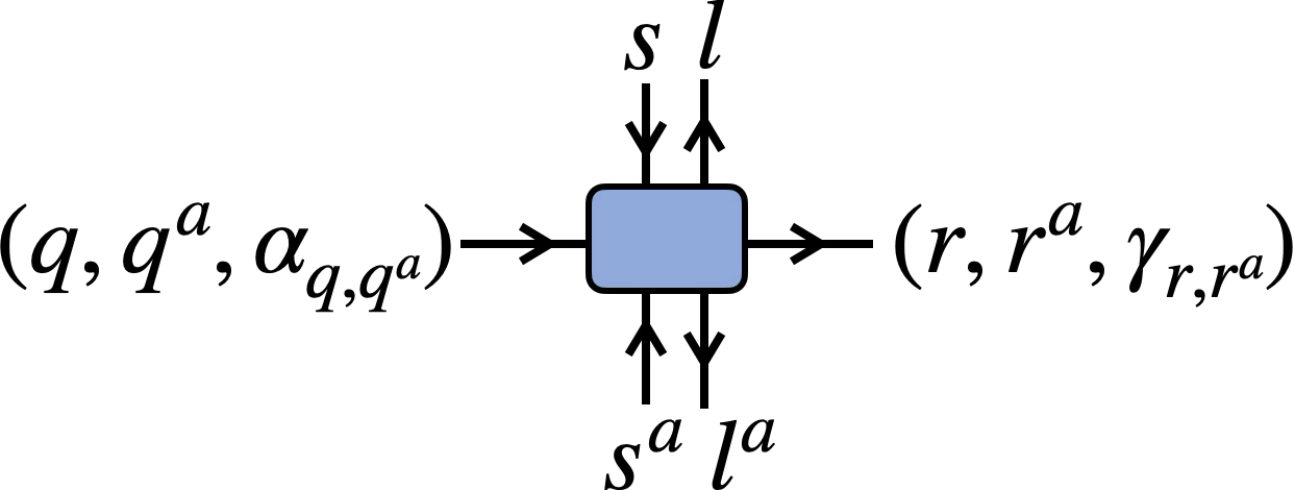}
    \caption{Schematic of the $U(1)$ gauge-invariant tensor for the Thermofield Double (TFD) state. This tensor extends the structure from Fig.~\ref{fig:GIMPS} by effectively containing two parallel copies of the pure-state tensor: one for the physical system (top indices, unprimed) and one for the auxiliary system (bottom indices, primed with $a$). Each copy independently enforces its own Gauss's law constraint and ensures the consistency between its virtual and physical link representations.
    }
    \label{fig:GIMPS_density}
\end{figure}

To enforce gauge invariance on the TFD state, we extend the pure-state symmetry structure to act on the doubled Hilbert space~\cite{Buyens:2016ecr}. This is achieved by applying the Gauss's law constraint independently to both the physical and auxiliary degrees of freedom. Following the same convention used for the pure state, we explicitly distinguish between the left virtual index $\alpha$ and the right virtual index $\gamma$ of each tensor $B_n$. These virtual indices are then augmented to carry separate charges for each space, $\alpha \rightarrow (q,q^a,\alpha_{q,q^a})$ and $\gamma \rightarrow (r,r^a,\gamma_{r,r^a})$, where $(q,r)$ belong to the physical space and $(q^a,r^a)$ to the auxiliary. The tensors decompose into a direct sum, $\left[B_n\right] =\oplus_{q,s,l,r; \,\, q^a,s^a,l^a,r^a} \left[B^{s, l; s^a,l^a}_n\right]_{\left(q,q^a,\alpha_{q,q^a}\right) ;\left(r,r^a,\gamma_{r,r^a}\right)}$, with the non-zero blocks defined as:
\begin{align}
    &\left[B^{s,l; s^a,l^a}_n\right]_{\left(q,q^a,\alpha_{q,q^a}\right) ;\left(r,r^a,\gamma_{r,r^a}\right)} \notag\\ 
    =&\left[b^{q,s,l,r; \,\, q^a,s^a,l^a,r^a}_n\right]_{\alpha_{q,q^a}, \gamma_{r,r^a}}\\
    & \cdot \delta_{l, q+\frac{s+(-1)^n}{2}}\, \delta_{r,l} \, \delta_{l^a, q^a+\frac{s^a+(-1)^n}{2}}\, \delta_{r^a,l^a} \,, \notag
\end{align}
where $\left[b^{q,s,l,r; \,\, q^a,s^a,l^a,r^a}_n\right] \in \mathbb{C}^{D_{q,q^a} \times D_{r,r^a}}$ and the total bond dimension of the MPS is given by $D=\sum_{q,q^a \in \mathbb{Z}} D_{q,q^a}$. The Kronecker deltas enforce the Gauss's law constraints for both the physical and auxiliary systems. Figure~\ref{fig:GIMPS_density} provides a schematic of this gauge-invariant tensor structure.

This maximally entangled structure at $\beta=0$ is captured by a simple MPS where the bond dimension is minimal ($D_{q,q^a}=1$ for all charge sectors) and the non-zero tensor elements locally enforce the mirroring between physical and auxiliary states:

\begin{align}
    \begin{split}
        \left[b^{q,s,l,r; \,\, q^a,s^a,l^a,r^a}_n\right]_{1,1} = \delta_{q,q^a} \delta_{s,s^a} \,,
    \end{split}
\end{align}

Finally, we determine the EoS, $P(\varepsilon)$. To ensure consistency with our non-equilibrium simulations, we use a system of the same size and mitigate boundary effects by averaging the local observables over the four central sites. For each inverse temperature $\beta$, we compute:
\begin{align}
    \begin{split}
        \varepsilon(\beta) &= \big\langle \langle\Phi(\beta)|\hat{T}^{tt}_n \otimes \hat{I}^a|\Phi(\beta)\rangle\big\rangle_n \\ 
    P(\beta) &= \big\langle \langle\Phi(\beta)|\hat{T}^{zz}_n \otimes \hat{I}^a|\Phi(\beta)\rangle\big\rangle_n,
    \end{split}
\end{align}
where $\big\langle \bullet \big\rangle_n$ stands for averaging over four sites at the center.
Mapping out this function, $P(\varepsilon)$, provides the equilibrium information required to analyze the system's non-equilibrium hydrodynamic behavior.

\section{Results and Discussion}\label{sec:results}
With the numerical framework ready, we move on to compute the thermodynamic properties, i.e., EoS, and real time evolution of a pure state. We visualize the spacetime propagation of the initial excitation, assess the emergence of hydrodynamic phenomena. Specifically, we characterize the microscopic dynamics of phase transition, which reveals distinct qualitative and quantitative signatures that distinguish the system's physical phases and transition behaviors.
In our numerical simulation, we take the grid size being $N=100$ with lattice spacing $a=0.5g^{-1}$. The electric field cut-off, bond dimension, step sizes in real or imaginary time evolution vary for different calculations with different purposes. We have ensured the convergency of all numerical results with respect to all these choices with details shown in the Appendix.

\subsection{Equation of State}
\begin{figure}[htbp]
    \centering
    \includegraphics[width=0.45\textwidth]{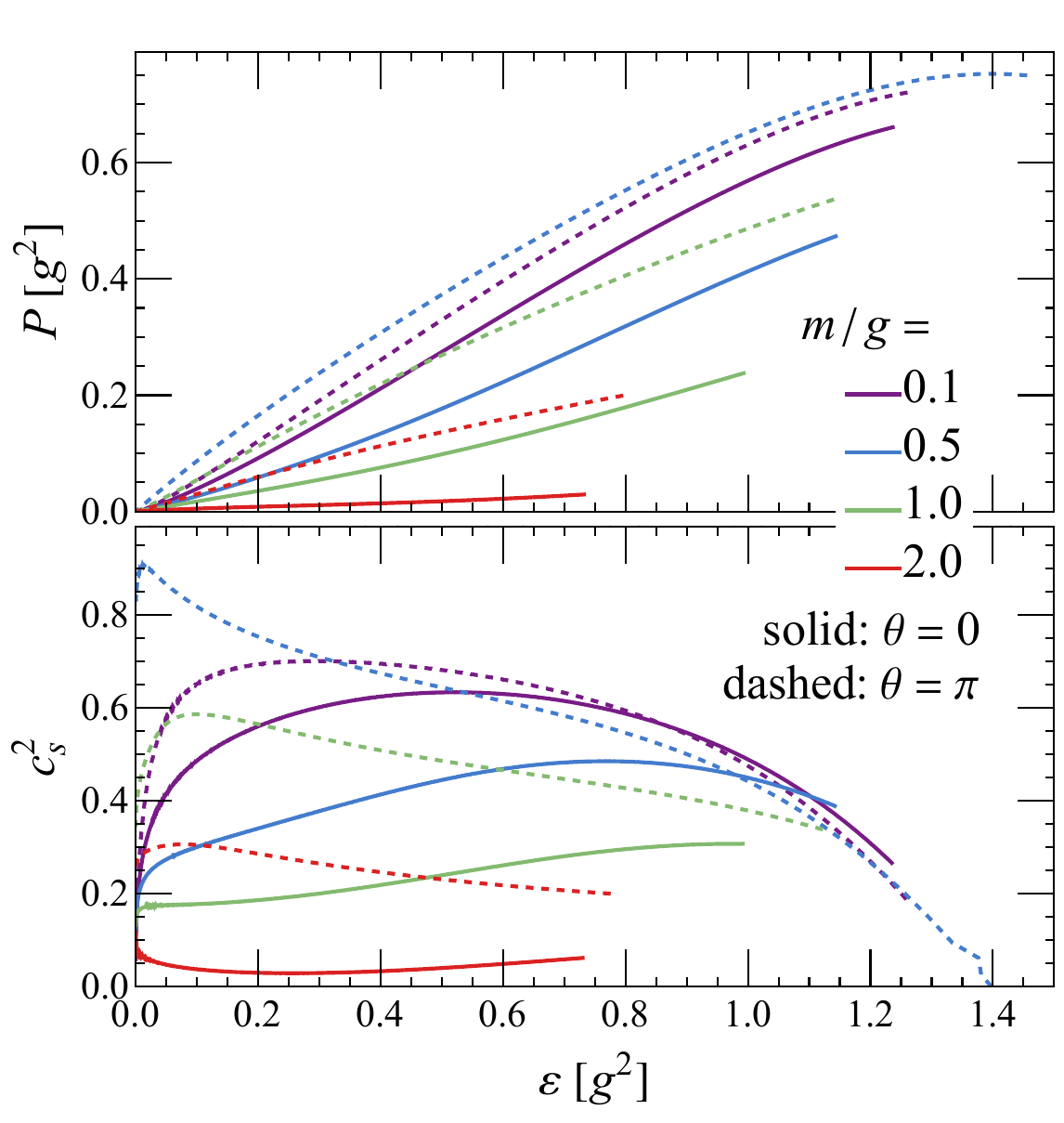}
    \caption{
        Pressure (upper) and speed of sound squared ($c_s^2$, lower)  versus energy density for the Schwinger model. Solid and dashed lines denote results for topological angles $\theta=0$ and $\theta=\pi$, respectively. The color scale maps to the mass-to-coupling ratio $m/g$, increasing from purple (low $m/g$) to red (high $m/g$).
    }
    \label{fig:eos}
\end{figure}

We show in Fig.~\ref{fig:eos} the thermodynamic pressure and the derived squared speed of sound, $c_s^2 = \mathrm{d}P/\mathrm{d}\varepsilon$, for the energy density of interest. At the topological angle $\theta=0$ (solid lines), the system exhibits a clear monotonic trend: for a given energy density $\varepsilon$, both pressure $P$ and $c_s^2$ decrease as the mass-to-coupling ratio $m/g$ increases. 
The reason is straightforward. The energy density contains two parts, the static one related to the rest mass, and the kinetic one from the thermal motion. A larger mass naturally corresponds to less thermal motion, and consequently less pressure.

This monotonic trend, however, breaks down at the topological angle $\theta=\pi$ (dashed lines). While the EoS still softens with increasing mass for $m/g > 0.5$, the trend conspicuously reverses at lower masses; specifically, the system at $m/g=0.5$ generates higher pressure and a greater speed of sound than at $m/g=0.1$.

We attribute this non-monotonicity to the well-known phase transition at $\theta=\pi$ and $m/g \approx 0.33$ associated with spontaneous symmetry breaking of parity~\cite{Hamer:1982mx, Schiller:1983sj, Byrnes:2002nv}. $m/g=0.1$ is in the parity-restored phase whereas $m/g=0.5$ is in the parity-broken phase in which the ground states become degenerate. This liberation of degrees of freedom stiffens the EoS, as the system can exert more pressure for a given energy density. The emergence of these new momentum carriers thus explains why the EoS stiffens as the mass crosses the critical point from the parity-restored side.

\begin{figure*}[htbp]\centering
    \includegraphics[width=0.85\textwidth]{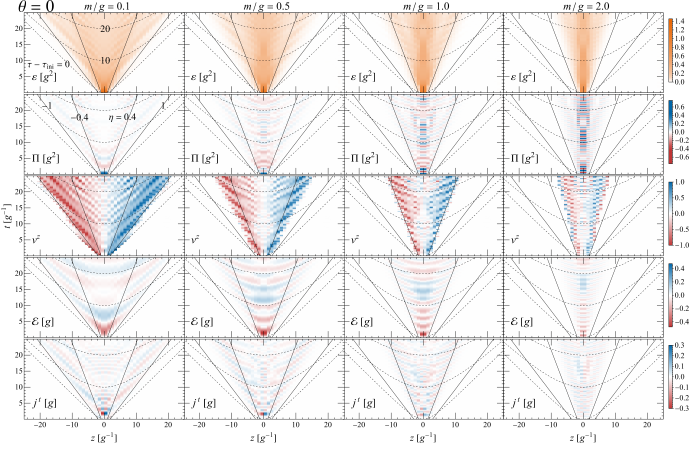}
    \includegraphics[width=0.85\textwidth]{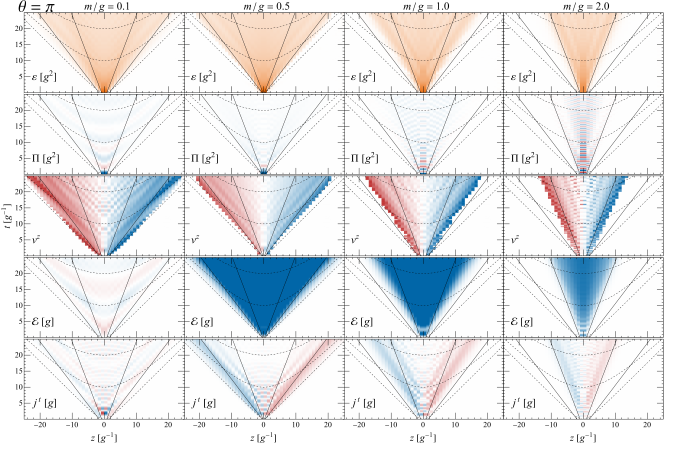}
    \caption{Spacetime evolution in the Schwinger model for several mass-to-coupling ratios $m/g \in \{0.1, 0.5, 1, 2\}$, with topological angle $\theta=0$ (first row) and $\theta=\pi$ (second row). Each subfigure contains five panels from top to bottom: energy density $\varepsilon$, bulk viscous pressure $\Pi$, flow velocity $v_z$, electric field strength $\mathcal{E}$, and charge density $j^t$, all as functions of spatial coordinate $z$ and time $t$. Color intensity represents the magnitude of each quantity. 
    In each subfigure, the black dashed lines indicate curves of constant proper time, with three lines (from bottom to top) corresponding to $\tau - \tau_\mathrm{ini} \in \{0, 10, 20\}$.  Black solid lines denote constant rapidity, with four lines (from left to right) indicating $\eta \in \{-1, -0.4, 0.4, 1\}$.
    As $m/g$ increases, both for $\theta=0$ and $\theta=\pi$, the evolution of $\varepsilon$, $\Pi$, and $v_z$ reveals a transition from fluid-like to non-fluid-like behavior. For $\theta = \pi$ and $m/g \in \{0.5,\, 1,\, 2\}$, the evolution of the electric field $\mathcal{E}$ and charge density $j^t$ exhibits characteristic features of phase transition behavior.}
    \label{fig:stress_tensor_evolution}
\end{figure*} 

\subsection{Hydrodynamic Behavior}
Fig.~\ref{fig:stress_tensor_evolution} displays the spacetime evolution of different observables at $\theta=0$ (first row) and $\theta=\pi$ (second row). In each row, from top to bottom are energy density ($\varepsilon$), bulk viscous pressure ($\Pi$), and flow velocity ($v_z$)--derived from the stress tensor according to~\eqref{eq:hydro_variables}, as well as electric field ($\mathcal{E}$) and charge density ($j^t$), respectively. As shown in Eq.~\eqref{eq:initial_state}, the evolution begins from a charge-neutral state with energy density localized at the central sites. This energy excitation rapidly expands and dynamically produces, within $t \sim 2\,g^{-1}$, a pair of positive and negative charges around the central region, which sets the stage for our subsequent dynamical analysis.

The evolution of the hydrodynamic variables ($\varepsilon, \Pi, v_z$) reveals a strong dependence on the mass-to-coupling ratio. With small fermion mass ($m/g=0.1$), the system's evolution closely resembles the Bjorken flow characteristic of relativistic heavy-ion collisions \cite{Bjorken:1982qr} for both $\theta=0$ and $\theta=\pi$: the medium spreads out within the lightcone, the velocity is roughly $v^z \approx z/t$. Particularly, the bulk pressure decreases in time and eventually vanishes, which is a signal that system approaches thermal equilibrium as noted before. The hydrodynamic behavior is more clear in the Milne space, which is discussed in details in our companion Letter~\cite{Shao_2025_short}. 

As $m/g$ increases, the deviation from Bjorken flow becomes more pronounced---the bulk pressure does not vanish, and the flow velocity is dominated by oscillations. 
Furthermore, the propagation speed of the excited region is mass-dependent. Larger $m/g$ ratios increase the particles' inertia, slowing their outward propagation. This effect is visible across all physical quantities in Fig.~\ref{fig:stress_tensor_evolution}: at any given time, systems with larger mass (e.g., $m/g=2$) exhibit a narrower excited region.
Notably, when mass is large, the velocity, electric field, and charge density profiles exhibit oscillation modes with both high and low frequencies, which was identified as the ``thumper'' mode in~\cite{Ikeda:2023vfk} reflecting the slow collective motion of high-frequency vibrating strings.

\subsection{Dynamical signatures of phase transition}
More interestingly, we observe in Fig.~\ref{fig:stress_tensor_evolution} that the spacetime profiles of the electric field ($\mathcal{E}$) and charge density ($j^t$) reveal two distinct dynamical regimes, corresponding to the parity-restored versus parity-spontaneously-broken phase transition.

The parity-restored phase occurs for all simulations at $\theta=0$ and for $m/g < 0.33$ at $\theta=\pi$ (e.g., $m/g=0.1$). In this phase, the initial charge pair propagates outwards, creating an unstable electric field string. The string's energy excites new particle-antiparticle pairs from the vacuum, and this repeated string-breaking process forms numerous secondary charge pairs that screen the primary ones. The final charge distribution, as seen in the evolution of the charge density $j^t$ (bottom panel), shows an alternating pattern of positive and negative charges. This complete screening effect neutralizes the electric field on a large scale, causing the electric field strength $\mathcal{E}$ in the central region to rapidly oscillate and decay to zero after a brief growth.

The parity-spontaneously-broken phase ($\theta=\pi, m/g > 0.33$) exhibits clearly different dynamics. Here, the background field stabilizes the electric string and strongly suppresses its breaking. The $j^t$ distributions (for $\theta=\pi, \, m/g \in \{0.5, 1, 2\}$) show that the primary charges propagate stably in opposite directions like nearly free particles, and they are connected by electric field that expands in space and does not decrease in time. Although the system still produces some secondary charges, they are far less than those in the parity-restored phase and quickly annihilate. For instance, in the case of $m/g=0.5$, for $t > 15g^{-1}$, the charge density in the central region has essentially decayed to zero. 

This is caused by the degenerate vacuum of the continuum theory, one has a vacuum electric field pointing to the left and the other is the opposite. The lattice theory introduces slight breaking of the parity symmetry and the latter is slightly more energetic.  
By measuring the expectation values of electric field ($\langle\mathcal{E}\rangle$) and electric field energy ($\langle\mathcal{E}^2\rangle$) for the blue areas in the plots, we found that the former tends to vanish while the latter remains the same as the vacuum.
Thus, the perturbed region has a positive electric field compared to the vacuum, while expansion of the blue area costs no energy.
These expectation values also indicate that the static electric field energy in the vacuum has converted to the (thermal) kinetic energy of gauge field particles, which is a distinct dynamical regime at finite temperature.
In this phase, one can separate a pair of charges to an arbitrary distance and create a uniform electric field string between them. It is also referred to as the deconfinement phase but for zero-temperature states~\cite{Tong, PintoBarros:2018bzz, Nandori:2010ij, Jentsch:2021trr}. 

\begin{figure}[htbp]
    \centering
    \includegraphics[width=0.48\textwidth]{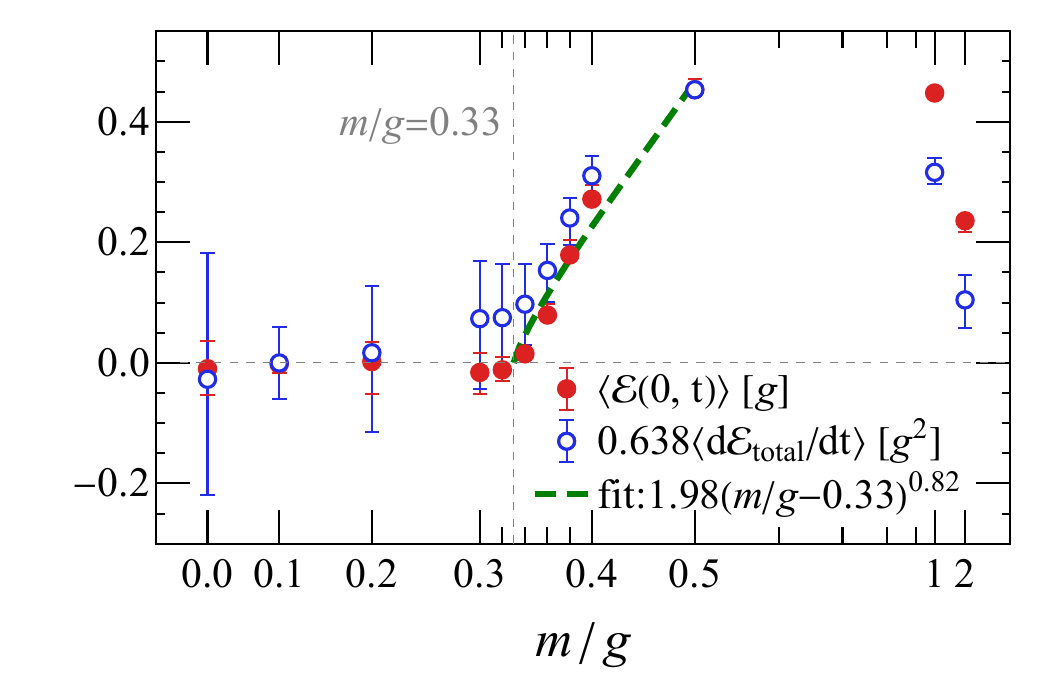}
    \caption{Two order parameters as a function of $m/g$ at the topological angle $\theta = \pi$. The plot shows the time-averaged central electric field, $\langle\mathcal{E}(0,t)\rangle$ (red solid line, filled circles), and the time-averaged growth rate of the total electric field, $\langle d\mathcal{E}_{\text{total}}/dt \rangle$ (blue solid line, open circles), scaled by a factor for clarity. Both averages are computed over the time interval $t = 10$--$25\,g^{-1}$. Error bars denote the full range of values (maximum and minimum) observed within this interval. The green dashed line is a power-law fit to the $\langle\mathcal{E}(0,t)\rangle$ data in the critical region.}
    \label{fig:order_param}
\end{figure}

To characterize the dynamical phase transition quantitatively, we estimate the electric field excitation in two ways. The first is the time-averaged electric field at the origin, $\langle\mathcal{E}(0,t)\rangle$, which directly probes the degree of local charge screening long after the initial excitation. The second is the time-averaged growth rate of the total electric field, $\langle d\mathcal{E}_{\text{total}}/dt \rangle$, which reflects the interplay between the expansion rate of the central field region and the overall strength of screening effects. The dependence of these parameters on $m/g$ is presented in Fig.~\ref{fig:order_param}. The close agreement between these distinct observables provides a robust picture of the underlying physics.

In the parity-restored phase ($m/g < 0.33$), both electric field estimators are negligible, a direct consequence of the complete charge screening from the string-breaking mechanism. In contrast, both observables increase sharply and simultaneously as the system crosses the critical point at $m/g \approx 0.33$. This behavior provides a clear dynamical signature of the phase transition.

It is noteworthy that deep in the parity-spontaneously-broken phase ($m/g \gtrsim 0.5$), the growth rate of the total electric field decreases as $m/g$ increases. This trend is consistent with our earlier observations in Fig.~\ref{fig:stress_tensor_evolution}. A larger mass increases particle inertia, which impedes the charge separation speed. This, in turn, slows the expansion of the constant-field region and reduces the growth rate of the total electric field.

Finally, we analyze the scaling of the central electric field to illustrate the system's critical behavior. The simulation results can be well described by a power-law fit of the form $f(x) = a(x - x_c)^b$ (green dashed line in Fig.~\ref{fig:order_param}), with $x_c = 0.33$ being the known location of the transition point. This indicates the scaling behavior near the phase transition. While this analysis highlights the critical nature of the transition, a rigorous determination of the exponent $b$ is beyond the scope of this work.

\section{Summary and Outlook}\label{sec:summary}
In this paper, we have investigated the real-time dynamics of the massive Schwinger model by invoking the Time-Evolving Block Decimation tensor network algorithm. We started from an initial state with a local energy excitation on top of the vacuum, which mimics the initial condition of a high-energy collision, and analyzed the space-time evolution of energy-momentum tensor, electric field, and charge density. Our analysis clarified the model's collective motion and phase structure from a dynamical perspective, revealing two key physical regimes.

First, examining the energy-momentum tensor reveals a transition in the system's collective behavior. For small mass-to-coupling ratio ($m/g \ll 1$), the system exhibits characteristics of Bjorken flow, indicating near-perfect relativistic hydrodynamics. As the ratio $m/g$ increases, the system's evolution progressively deviates from this hydrodynamic description. Whether a small mass or a strong coupling is the most important for the onset of hydrodynamics requires further investigations with other field theory models.

Second, our analysis of the electric field and charge density reveals the phase transition that spontaneously breaks parity, from a dynamical perspective. In the parity-restored phase ($\theta=0$, or $m/g < 0.33$ at $\theta=\pi$), the string-breaking mechanism dominates, rapidly producing particle-antiparticle pairs that screen the charge and neutralize the large-scale electric field. In contrast, the parity-spontaneously-broken phase ($\theta=\pi$, $m/g > 0.33$) is characterized by the existence of arbitrarily long strings. Here, the primary charges propagate like nearly free particles, creating an expanding region of stable, nearly constant electric field between them. 

We quantify this distinction at $\theta=\pi$ with two order parameters: the late-time central electric field and the growth rate of the total electric field. Both observables are consistent with zero in the parity-restored phase and increase sharply at the critical point ($m/g \approx 0.33$), signaling the transition. A power-law fit confirms that the behavior near this point exhibits the scaling expected of a critical phase transition.

Our space-time evolution of the electric field at the parity-spontaneously-broken phase seems to resemble what one would expect in the deconfinement phase. The ``confinement and deconfinement'' phase transition in the Schwinger model has been widely discussed in the literature, and further realized in cold atom experiments (see e.g.~\cite{Martinez:2016yna, Surace:2019dtp, Zhou:2021kdl, Cheng:2022jnw, Mildenberger:2022jqr, Xiang:2025qhq}) which truncate the electric field up to two states. Studying the property of confinement with low electric field truncation and in a finite temperature system needs further careful investigations. 

\vspace{5mm}
\section*{Acknowledgments}
We thank Sangyong Jeon, Weiyao Ke, Tongyu Lin, Dmitri Kharzeev, Li Yan, Yi Yin, Yuanchen Zhao, and Pengfei Zhuang for discussions. This work is supported by NSFC under grant No. 12575143 and Tsinghua University under grants No. 04200500123, No. 531205006, and No. 533305009.
Computations are performed at Center of High performance computing, Tsinghua University.

\bibliography{ref}

\begin{appendix}
\begin{widetext}
\section{Calculation Parameters and Convergence Tests}
\begin{figure*}[htbp]
    \centering
    \includegraphics[width=1\textwidth]{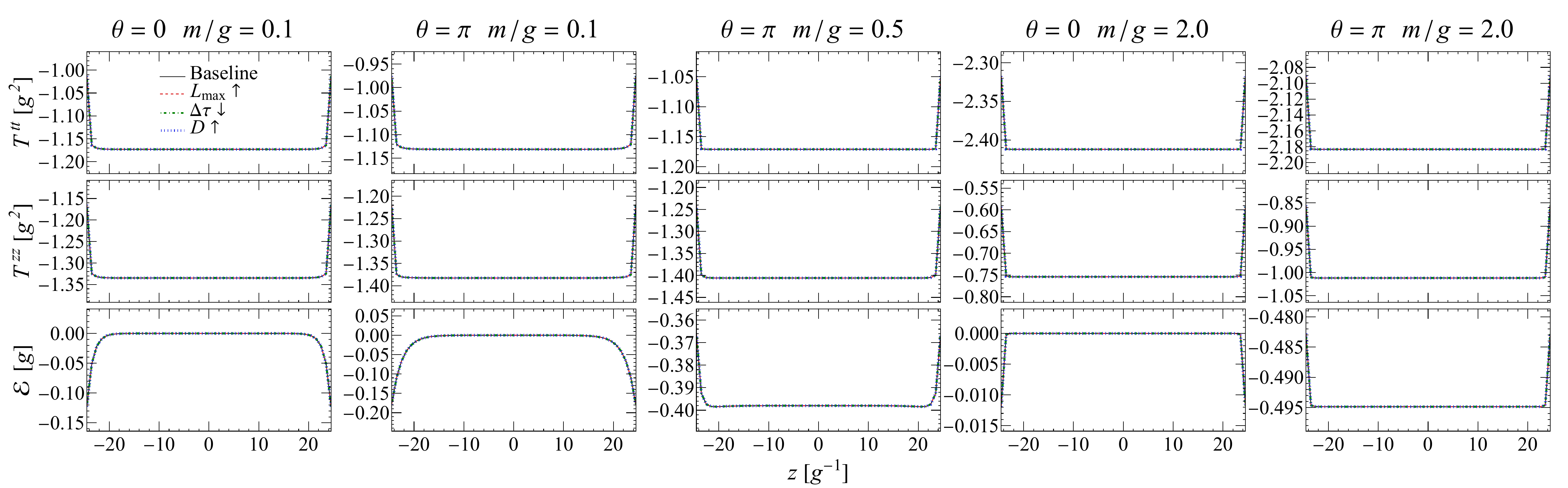}
    \caption{
    \textbf{Convergence of ground state calculations.}
    Spatial profiles of the energy-momentum tensor components ($T^{tt}$, $T^{zz}$) and the electric field ($\mathcal{E}$) for representative cases.
    \textbf{Convergence:} The baseline calculation (black solid) is shown to be converged, as it perfectly overlaps with results from individually varied parameters: an increased cutoff $L_{\text{max}}$ (red dashed), a smaller time step $\Delta\tau$ (green dot-dashed), and a larger bond dimension $D$ (blue dotted).
    \textbf{Parameters:} Baseline is $\{L_{\text{max}}=3, \Delta\tau=0.01/g, D=100\}$. Varied parameters are $\{L_{\text{max}}=4\}$, $\{\Delta\tau=0.004/g\}$, and $\{D=200\}$.
    }
    \label{fig:GSconvergence}
\end{figure*}

\begin{figure*}[htbp]
    \centering
    \includegraphics[width=1\textwidth]{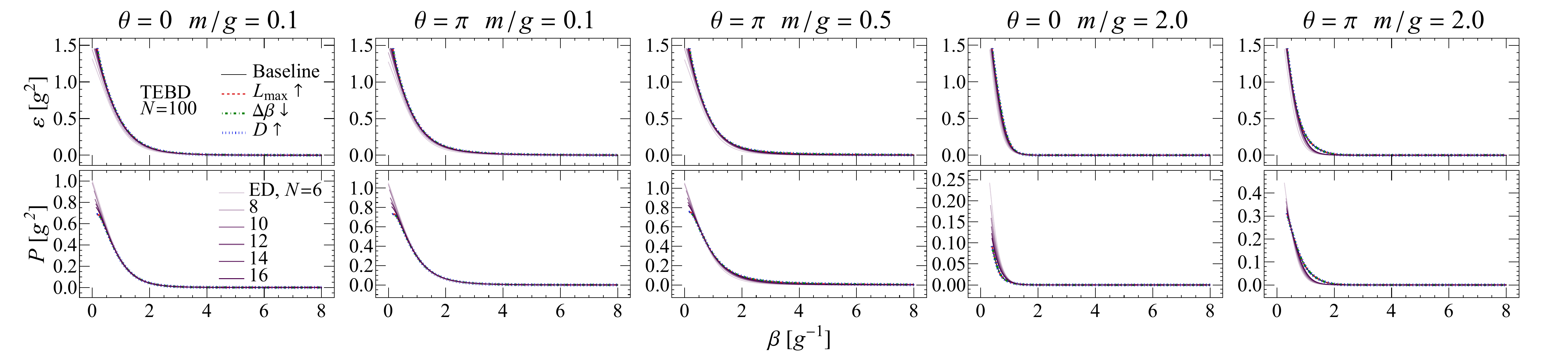}
    \caption{\textbf{EoS convergence and accuracy benchmark.}
    The energy density $\varepsilon$ and pressure $P$ versus inverse temperature $\beta$.
    \textbf{Convergence:} The baseline calculation (black solid) is shown to be converged, as it perfectly overlaps with results from varied parameters: increased $L_{\text{max}}$ (red dashed), smaller step $\Delta\beta$ (green dot-dashed), and larger $D$ (blue dotted).
    \textbf{Accuracy:} Our baseline result also shows strong agreement with exact diagonalization (purple solid lines, for $N=6$ to $16$).
    \textbf{Parameters:} Baseline is $\{L_{\text{max}}=8, \Delta\beta=0.002/g, D=100\}$ for $\beta \le 3/g$ and $\{L_{\text{max}}=3, \Delta\beta=0.02/g, D=100\}$ for $\beta > 3/g$. Varied parameters are $\{L_{\text{max}}=10 \text{ or } 5\}$, $\{\Delta\beta=0.0008/g \text{ or } 0.008/g\}$, and $\{D=200\}$.
    }
    \label{fig:EOSconvergence}
\end{figure*}

\begin{figure*}[htbp]
    \centering
    \includegraphics[width=1\textwidth]{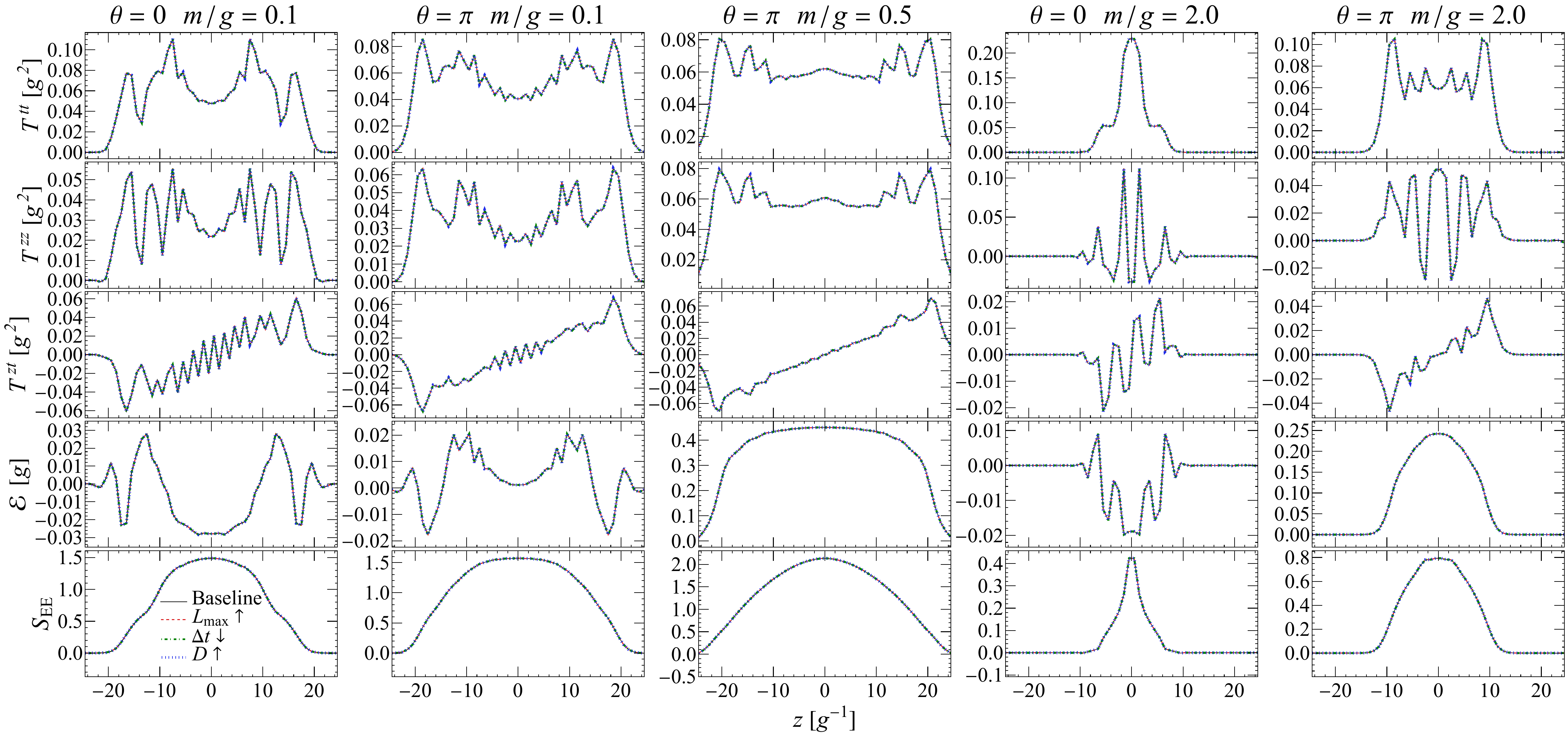}
    \caption{
    \textbf{Convergence of real-time evolution.}
    Spatial profiles of key quantities at the final evolution time, $t=25/g$.
    \textbf{Convergence:} The baseline calculation (black solid) is shown to be converged, as it perfectly overlaps with results from individually varied parameters: an increased cutoff $L_{\text{max}}$ (red dashed), a smaller time step $\Delta t$ (green dot-dashed), and a larger bond dimension $D$ (blue dotted).
    \textbf{Parameters:} Baseline parameters are optimized for each mass ratio, using $\{L_{\text{max}}=3, \Delta t=0.004/g\}$ with bond dimension $D=250$ for $m/g=0.1, 2.0$ and $D=500$ for $m/g=0.5$. The varied parameters tested are $\{L_{\text{max}}=4\}$, $\{\Delta t=0.001/g\}$, and $\{D=500 \text{ or } 750\}$.
    }
    \label{fig:RTEconvergence}
\end{figure*}
\end{widetext}

This appendix establishes the numerical robustness of our simulations by verifying convergence with respect to both algorithmic parameters and the lattice discretization. We first demonstrate convergence with respect to the core algorithmic parameters---including the electric field cutoff ($L_{\text{max}}$), time step, and bond dimension ($D$)---for our ground state preparation, EoS calculations, and real-time evolution. These tests were performed on representative parameter sets, $\{\theta=0, m/g=0.1, 2.0\}$ and $\{\theta=\pi, m/g=0.1, 0.5, 2.0\}$, chosen to cover extremal mass ratios and to reliably probe both sides of the phase transition at $\theta=\pi$. We then confirm that lattice discretization effects are negligible, as the simulated dynamics stabilize for systems with lattice size $N \ge 100$. The fully converged parameters used for our production runs are summarized in Table~\ref{tab:simulation_parameters}.

\begin{table}[htbp]
\centering
\caption{Parameters that ensure convergence in the numerical simulations. The table lists the electric field cutoff ($L_{\text{max}}$), the imaginary and real time step sizes ($\Delta\tau$, $\Delta\beta$, and $\Delta t$), and the maximum bond dimension ($D$). Common parameters for all simulations are a lattice of size $N=100$ with spacing $a=1/(2g)$ and an initial excitation strength of $\varphi = \pi$.}
\label{tab:simulation_parameters}
\begin{tabular}{lccc}
\hline
\textbf{Simulation Stage} & $L_{\text{max}}$ & Time step & $D$ \\
\hline
Ground state preparation                  & 3   & $\Delta\tau = 0.01/g$          & 100 \\
Real-time evolution                       & 3   & $\Delta t = 0.004/g$           & 500 \\
EoS calculation: $\beta \le 3/g$         & 8   & $\Delta\beta = 0.002/g$         & 100 \\
EoS calculation: $\beta > 3/g$  & 3   & $\Delta\beta = 0.02/g$          & 100 \\
\hline
\end{tabular}
\end{table}

We begin by verifying the convergence of our ground state calculations. Figure~\ref{fig:GSconvergence} displays the spatial profiles of the key observables: the energy-momentum tensor components $T^{tt}$ and $T^{zz}$, and the electric field $\mathcal{E}$. The charge density $j^t$ is not shown as its convergence follows from $\mathcal{E}$ via Gauss's Law, and $T^{zt}$ is omitted as it is identically zero in the ground state. As detailed in the caption and shown in the figure, the perfect overlap of the curves confirms our ground state calculations are well-converged.

Next, we validate our EoS calculations in Fig.~\ref{fig:EOSconvergence}. This test confirms both numerical convergence against algorithmic parameters and physical accuracy against exact diagonalization. We plot the energy density $\varepsilon$ and pressure $P$ versus inverse temperature $\beta$. The excellent agreement with exact diagonalization benchmarks our method's accuracy, while the stability against parameter variations confirms its convergence.

Then, we demonstrate the convergence of our real-time evolution simulations. We test convergence at the final time, $t=25/g$, as this is the most computationally demanding point where numerical errors and entanglement have maximally accumulated. Figure~\ref{fig:RTEconvergence} shows the spatial profiles of key quantities, including the energy-momentum tensor components ($T^{tt}, T^{zz}, T^{zt}$), the electric field $\mathcal{E}$, and the entanglement entropy $S_{\text{EE}}$. The latter is particularly important for time evolution, as entanglement growth often limits the duration of simulations. The perfect overlap of the curves confirms that our results are robust against variations in the numerical parameters.

\begin{figure}[!htbp]
    \centering
    \includegraphics[width=0.45\textwidth]{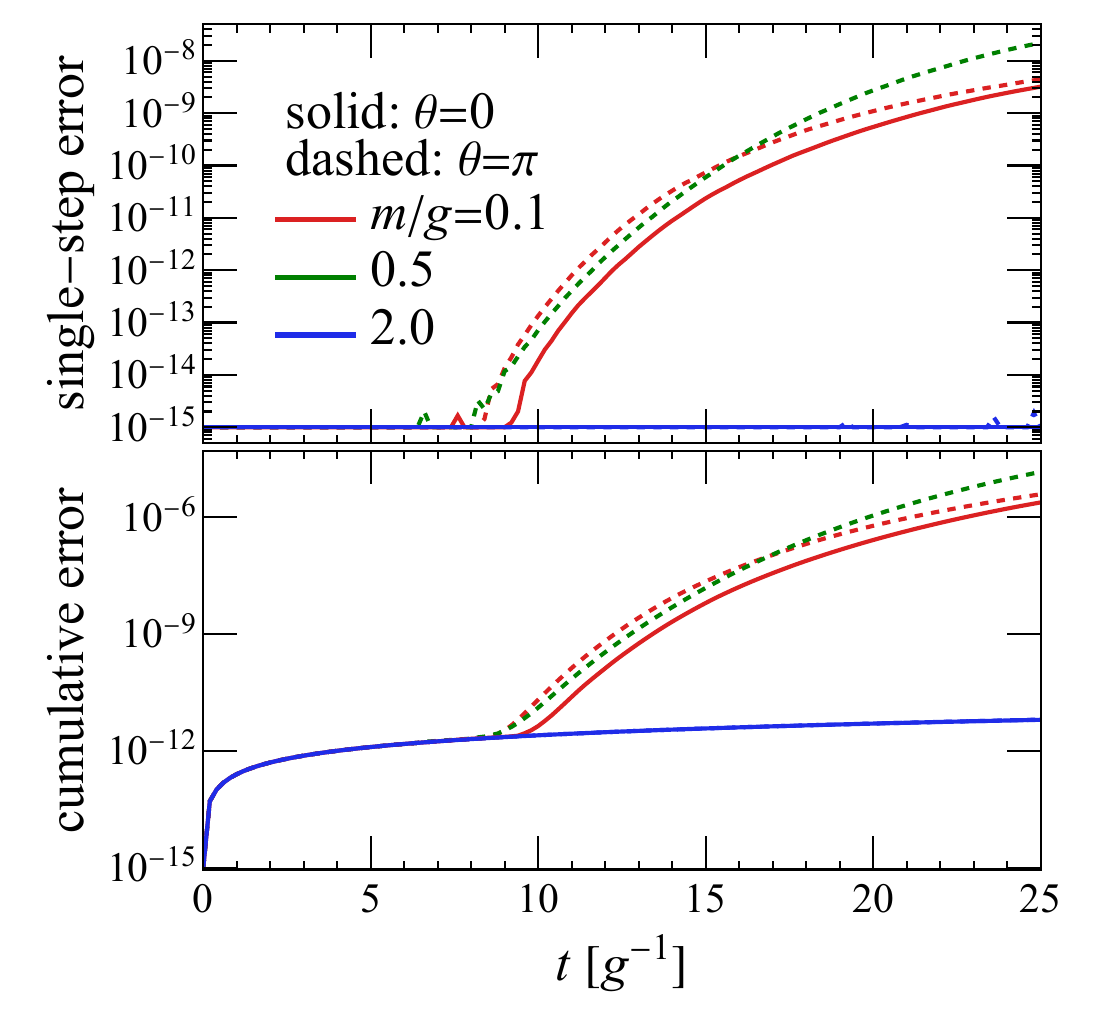}
    \caption{
    \textbf{TEBD truncation errors during time evolution.} 
    Single-step (top) and cumulative (bottom) SVD errors for $\theta=0$ (solid lines) and $\theta=\pi$ (dashed lines). 
    Colors correspond to $m/g=0.1$ (red), $0.5$ (green), and $2.0$ (blue).
    A floor value of $10^{-15}$ is used for plotting, chosen to be just above the machine epsilon ($\approx 2.22 \times 10^{-16}$) of the double-precision floating-point number employed. The small magnitude of the errors confirms the high precision of the simulations.
    }
    \label{fig:TruncError}
\end{figure}

To further quantify this, we perform two stringent checks. First, we monitor the singular value decomposition (SVD) truncation error, with the single-step and cumulative errors plotted in Fig.~\ref{fig:TruncError}. For the most demanding case ($\theta=\pi, m/g=0.5$), the maximum single-step error was $2.21 \times 10^{-8}$ and the total accumulated error remained below $1.44 \times 10^{-5}$. 
Second, we checked energy conservation and found the maximum relative change between the initial and final total energy, $|E_{\text{final}}-E_{\text{initial}}|/|E_{\text{initial}}|$, to be $1.73 \times 10^{-5}$, also for the $\theta=\pi, m/g=0.5$ case. Together, these checks confirm that our simulation parameters are sufficient to achieve high-precision results.

\begin{figure}[!htbp]
    \centering
    \includegraphics[width=0.5\textwidth]{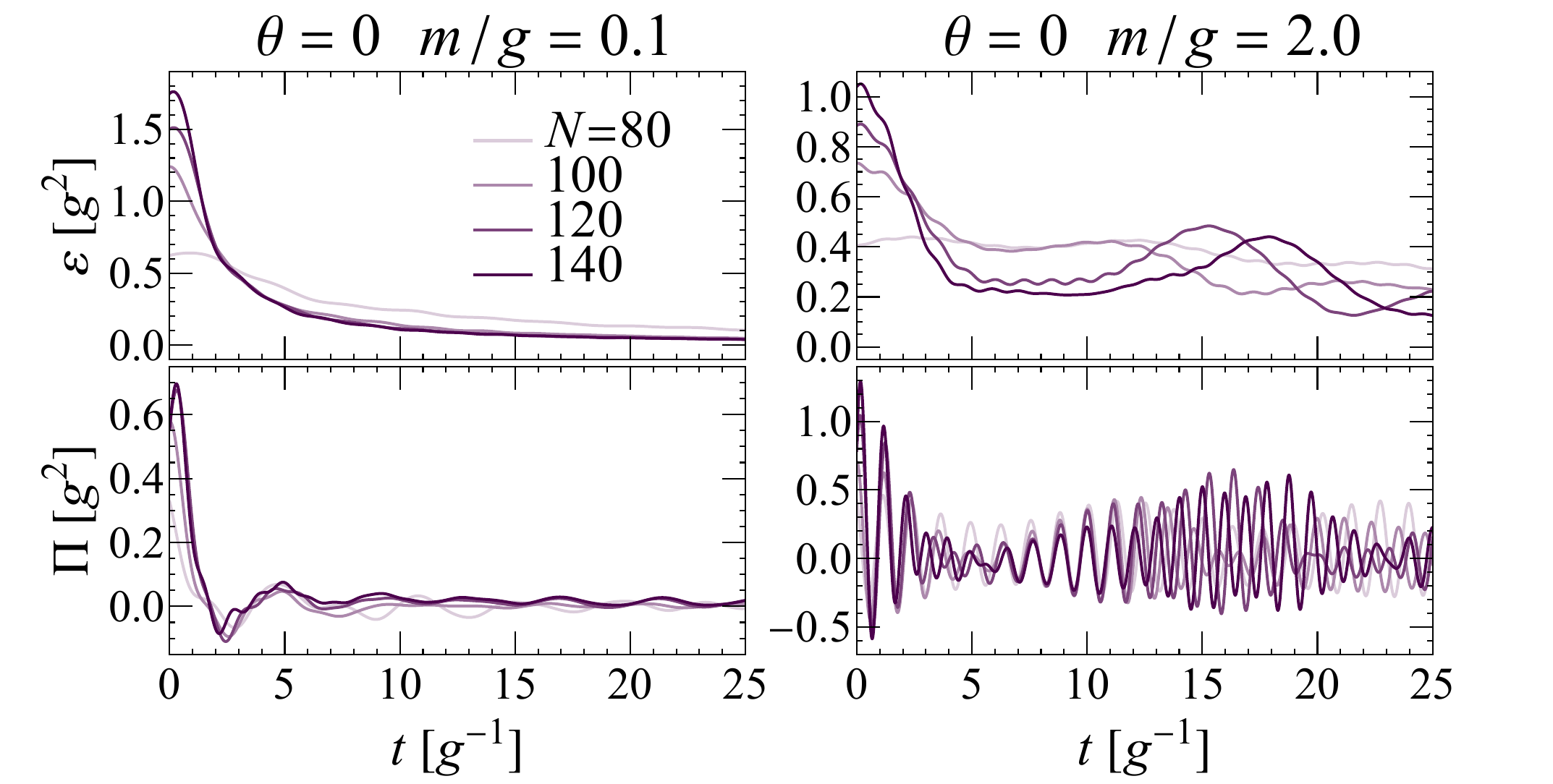}
    \caption{\textbf{Lattice independence check.}
    Evolution of the energy density $\varepsilon$ (top) and bulk pressure $\Pi$ (bottom) in the central rapidity region. Different curves correspond to lattice sizes $N=80, 100, 120, 140$ (darker shades for larger $N$). For both coupling ratios, the results for $N \ge 100$ are consistent: the late-time curves converge for $m/g=0.1$ (signaling hydrodynamization), and show similar qualitative behavior for $m/g=2.0$ (no hydrodynamization).
    }
    \label{fig:ContinueLimit}
\end{figure}
In addition to algorithmic convergence, we tested the robustness of our results against lattice discretization artifacts. To this end, we evolved the system on lattices of various sizes ($N=80$, $100$, $120$, $140$) for representative cases ($\theta=0$ with $m/g=0.1$, $2.0$), with keeping a constant total volume ($N\,a=50\, g^{-1}$). To ensure a fair comparison focused on the dynamics, the total initial energy of the system was held constant across all lattice sizes. This required adjusting the initial excitation protocol for each case, as our benchmark $N=100$ setup could not produce the target energy on other lattices. Our standard protocol for $N=100$ uses an excitation strength of $\varphi=\pi$ on sites $\{50, 51\}$. To achieve the same total energy, the protocols for other lattice sizes were adjusted as follows. For $N=120$ and $N=140$, the strength $\varphi$ was modified on two central sites: for $N=120$, $\varphi=0.415\pi$ when $m/g=0.1$ and $0.380\pi$ when $m/g=2.0$; for $N=140$, $\varphi=0.365\pi$ when $m/g=0.1$ and $0.323\pi$ when $m/g=2.0$. The $N=80$ case required a wider four-site excitation on sites $\{39, 40, 41, 42\}$ with $\varphi=0.825\pi$ when $m/g=0.1$ and $0.790\pi$ at $m/g=2.0$.

As shown in Fig.~\ref{fig:ContinueLimit}, the dynamics for lattice sizes $N \ge 100$ stabilize and become consistent. Specifically, for the hydrodynamizing case ($m/g=0.1$), the curves for $\varepsilon$ and $\Pi$ show clear quantitative convergence at late times. For the non-hydrodynamizing case ($m/g=2.0$), while full quantitative convergence is not observed, the results for $N=100$, $120$, and $140$ exhibit the same distinct qualitative behavior, setting them apart from the $N=80$ simulation. For the case that $m/g=0.1$ where hydrodynamic behavior has been observed, the stabilization in both $\varepsilon$ and $\Pi$ gives us confidence that our choice of $N=100$ for the main simulations (see Table~\ref{tab:simulation_parameters}) is sufficient to reliably capture the underlying physics, without significant discretization effects at this energy scale. While curves in the $m/g=2.0$ case do not exhibit the trend of convergence, they all show very sizable and non-vanishing quantum oscillations in the bulk pressure, which makes the conclusion of the absence of hydrodynamic behavior robust.

\end{appendix}

\end{document}